\documentclass[prc,twocolumn,showpacs,showkeys,nofootinbib,superscriptaddress]{revtex4}
\usepackage{graphicx}
\usepackage{dcolumn}
\usepackage{epsfig}
\usepackage{bm}

\begin{document}

\topmargin -0.50in

\title{Nuclear and particle physics aspects of the $2\nu\beta\beta$-decay of $^{150}Nd$}

\author{Rastislav Dvornick\'y}

\affiliation{Department of Nuclear
Physics, Comenius University, Mlynsk\'a dolina F1, SK--842 15
Bratislava, Slovakia}
\author{Fedor \v Simkovic}

\altaffiliation{On  leave of absence from Department of Nuclear
Physics, Comenius University, Mlynsk\'a dolina F1, SK--842 15
Bratislava, Slovakia}
\affiliation{Institute f\"{u}r Theoretische Physik der Universit\"{a}t
T\"{u}bingen, D-72076 T\"{u}bingen, Germany}
\author{Amand Faessler}

\affiliation{Institute f\"{u}r Theoretische Physik der Universit\"{a}t
T\"{u}bingen, D-72076 T\"{u}bingen, Germany}



\begin{abstract}
A discussion is given on possible realization of the Single State Dominance (SSD)
hypothesis in the case of the two-neutrino double beta decay ($2\nu\beta\beta$-decay)
of $^{150}Nd$ with $1^-$ ground state of the intermediate nucleus. We conclude that
the SSD hypothesis is expected to be ruled out by precision measurement of 
differential characteristics of this process in running NEMO 3 or planed SuperNEMO 
experiments unlike some unknown low-lying $1^+$ state of $^{150}Pm$ does exist. This 
problem can be solved via (d,$^{2}He$) charge-exchange  experiment on $^{150}Sm$. 
Further, we address the question about possible violation of the Pauli exclusion 
principle for neutrinos and its consequences for the energy distributions of the
$2\nu\beta\beta$-decay of $^{150}Nd$. This phenomenon might be a subject of interest 
of NEMO 3 and SuperNEMO experiments as well. 
\end{abstract}

\maketitle


\section{Introduction}

The main interest in the double beta decay is connected with the neutrinoless mode
($0\nu\beta\beta$-decay) as a probe for physics beyond the Standard Model (SM) of 
electroweak interactions. The detection of  double beta decay with emission of
two neutrinos ($2\nu\beta\beta$-decay), which is an allowed process of second order
in the SM, provides the possibility for experimental determination of the 
corresponding nuclear matrix elements (NME's). At present, $2\nu\beta\beta$-decay 
has been measured for ten nuclei \cite{barab}.

\begin{table*}[t]
\squeezetable
\renewcommand{\arraystretch}{1.9}    
\caption{Basic characteristics of the electron capture (EC) and $\beta^-$-decay of the 
ground state of the intermediate nucleus for $A=100$ and $150$ systems \cite{zdes}. 
$^\dagger$The half-life was estimated by assuming that values of
the $EC$ and $\beta^-$ matrix elements are comparable for a given isotope. \\
} 
 \begin{tabular}{lcccccc}\hline\hline
Type &Transition & Q [MeV] & $J^\pi$ & $T^{exp}_{1/2}$ & 
$\left|{\cal M}(J^\pi)\right|$ & $log~ ft$\\
\hline
$\beta^-$ & 
${^{100}Tc}(1^+_{g.s.}) \rightarrow {^{100}Ru}(0^+_{g.s.})$ & 
3.202 & $1^+$ & $15.8~ [s]$& 0.69 & 4.6\\ 
$EC$ & 
$ {^{100}Tc}(1^+_{g.s.}) \rightarrow {^{100}Mo}(0^+_{g.s.})$ &
0.168 & $1^+$ & $8.77~10^5~[s]$ & 0.82 & 4.5 \\
$\beta^-$ & 
${^{150}Pm}(1^-_{g.s.}) \rightarrow {^{150}Nd}(0^+_{g.s.})$ &
0.086 & $1^-$ & $\approx 6~10^{16}~[y]^\dagger$ & $\approx 0.016^\dagger$ & 8.0\\
$EC$ & 
${^{150}Pm}(1^-_{g.s.}) \rightarrow {^{150}Sm}(0^+_{g.s.})$ &
3.454 & $1^-$ & $9.64~ 10^3~ [s]$ & 0.016 & 7.9 \\ 
\hline\hline
\label{tab.pol}
\end{tabular}
\end{table*}

A subject of interest is the single state dominance (SSD) hypothesis proposed by Abad
et al. \cite{aba84} some times ago. It was suggested that $2\nu\beta\beta$-decays
with $1^+$ ground state of the intermediate nucleus (e.g., A=100, 116 and 128
nuclear systems) are solely determined by the two virtual $\beta$-transitions:
(i) the first one connecting the ground state of the initial nucleus  with
$1^+_1$ intermediate state; (ii) the second one proceeding from the $1^+_1$ state
to the final ground state. An alternative higher state dominance (HSD) approach
assumes that high-lying energy states of the intermediate nucleus give the main
contribution to the $2\nu\beta\beta$-decay NMEs. In this case the sum of the two
lepton energies in the denominators of the NMEs can be replaced with their average
value \cite{sim01}. 

Recently, it has been pointed out that the $2\nu\beta\beta$-decay allows to
investigate also particle properties, in particular whether the Pauli
exclusion principle is violated for neutrinos, and thus, neutrinos obey
at least partly the Bose-Einstein statistics \cite{dosm,Barabash:2007gb}. 
For the $2\nu\beta\beta$-decay of $^{76}Ge$ and $^{100}Mo$ 
results of numerical calculations of the total rates and various distributions 
were presented in \cite{Barabash:2007gb}.

Currently, much attention is paid to the double beta decay of $^{150}Nd$
due to a large Q-value  and low background from natural radioactivity.
This isotope  is considered to be a proper candidate for planed SuperNEMO, 
DCBA(Drift Chamber Beta-ray Analyzer) and SNO+ experiments. 
In this contribution two problems are addressed: Can the SSD hypothesis be
confirmed or ruled out for the $2\nu\beta\beta$-decay of $^{150}Nd$? 
Is it possible to study also the violation of the Pauli exclusion principle
for neutrinos in this process?

\vspace{0.6cm}

\section{The SSD versus HSD study}

Till now the issue of the SSD hypothesis has been not addressed in the case of the
$2\nu\beta\beta$-decay of $^{150}Nd$ as the ground state of the intermediate nucleus
is $1^-$ state. It looks simply unlikely that the SSD is realized through forbidden EC
and $\beta^-$-decay, nevertheless,  this issue has not been checked yet.  
However, a negative energy difference between the initial $^{150}Nd$ and the intermediate 
$^{150}Pm$ ground states favors this nuclear system for the SSD/HSD analysis 
via differential characteristics.

It is interesting to compare the $2\nu\beta\beta$-decays of $^{100}Mo$ and $^{150}Nd$
in respect to the SSD/HSD hypothesis. The basic characteristics of these two nuclear
systems are given in Table \ref{tab.pol}. The presented nuclear matrix elements 
for EC of $^{100}Tc$ and $\beta$-decays of $^{100}Tc$ and $^{150}Pm$ were calculated
from experimental half-lives (see Table \ref{tab.pol}). We have
\begin{eqnarray}
\left[{T^{\beta, EC}_{1/2}(J^{\pi_i}_i \rightarrow  0^{+}_f )}\right]^{-1} =
~~~~~~~~~~~~~~
\nonumber\\
\frac{m_e}{6 \pi^3 \ln{2}}~ (G_{\beta} m^2_e)^{{2}} \left|{\cal M}(J^\pi)\right|^2
f_{\beta, EC} (Z,E_i-E_f).
\end{eqnarray}
The phase space integrals are given by \newpage
\begin{eqnarray}
f_\beta (Z,E_i-E_f) = 
~~~~~~~~~~~~~~~~~~~~
\nonumber\\
\frac{1}{m^5_e}
\int_{m_e}^{E_i-E_f} F_0(Z,p_{0}) p p_{0} (E_i - E_f - p_{0})^2 dp_{0}, 
\end{eqnarray}
\begin{eqnarray}
f_{EC}(Z,E_i-E_f) = 2\pi^2  
\left(\frac{1}{m^3_e} \frac{Z^3}{\pi a^3_e}\right)  
\frac{(E_i - E_f + \varepsilon_b)^2}{m^2_e}. 
\end{eqnarray}
Nuclear matrix elements can be written as 
\begin{equation}
{\cal M}(J^\pi) = <0^{+}_f||{\cal O}(J^{\pi})||J^{\pi_i}_i> 
\end{equation}
with
\begin{eqnarray}
{\cal O}_k (1^+) = i g_A \sum_m \tau^+_m \left(\vec\sigma_m\right)_k,
\end{eqnarray}
\begin{eqnarray}
{\cal O}_k (1^-) = \left(\frac{\alpha Z}{2}\right)
\sum_m \tau^+_m \frac{1}{R}\left(\vec{x_m} - 
i g_A \vec{x_m}\times\vec\sigma_m\right)_k.
\label{eq.op} 
\end{eqnarray}
Here, $J^{\pi}=1^{+},1^{-}$. $E_i$, $E_f$ are energies of the initial and final 
nuclei, respectively. $g_A$ is the axial-vector coupling constant ($g_A=1.254$). 
R is nuclear radius.

Further we assume that the nuclear matrix element for EC of 
the ground state $^{150}Pm$ is comparable with that for 
$\beta^-$-decay of this isotope. Both matrix elements are suppressed
by the same coulombic factor ($\alpha Z/2$) and a similar situation
occurs also for A=100 system (see Table \ref{tab.pol}) \cite{doi85}. 
Under this assumption we find the half-life
of the EC of $^{150}Pm$ to be about $6~10^{16}$ years, i.e., not measurable
with help of current technologies. The SSD prediction for the $2\nu\beta\beta$-decay 
half-life is $4~10^{24}$ years in complete disagreement with the experimental 
value $9.7~10^{18}$ years \cite{barab}. We recall that for the 
$2\nu\beta\beta$-decay of $^{100}Mo$ there is a rather good agreement 
between the SSD calculated value $6.8~10^{18}$ years  and the measured value 
$7.1~10^{18}$ years. 

\begin{figure*}[t]
    \epsfxsize=0.9\textwidth
    \epsfysize=0.4\textwidth
    \epsffile{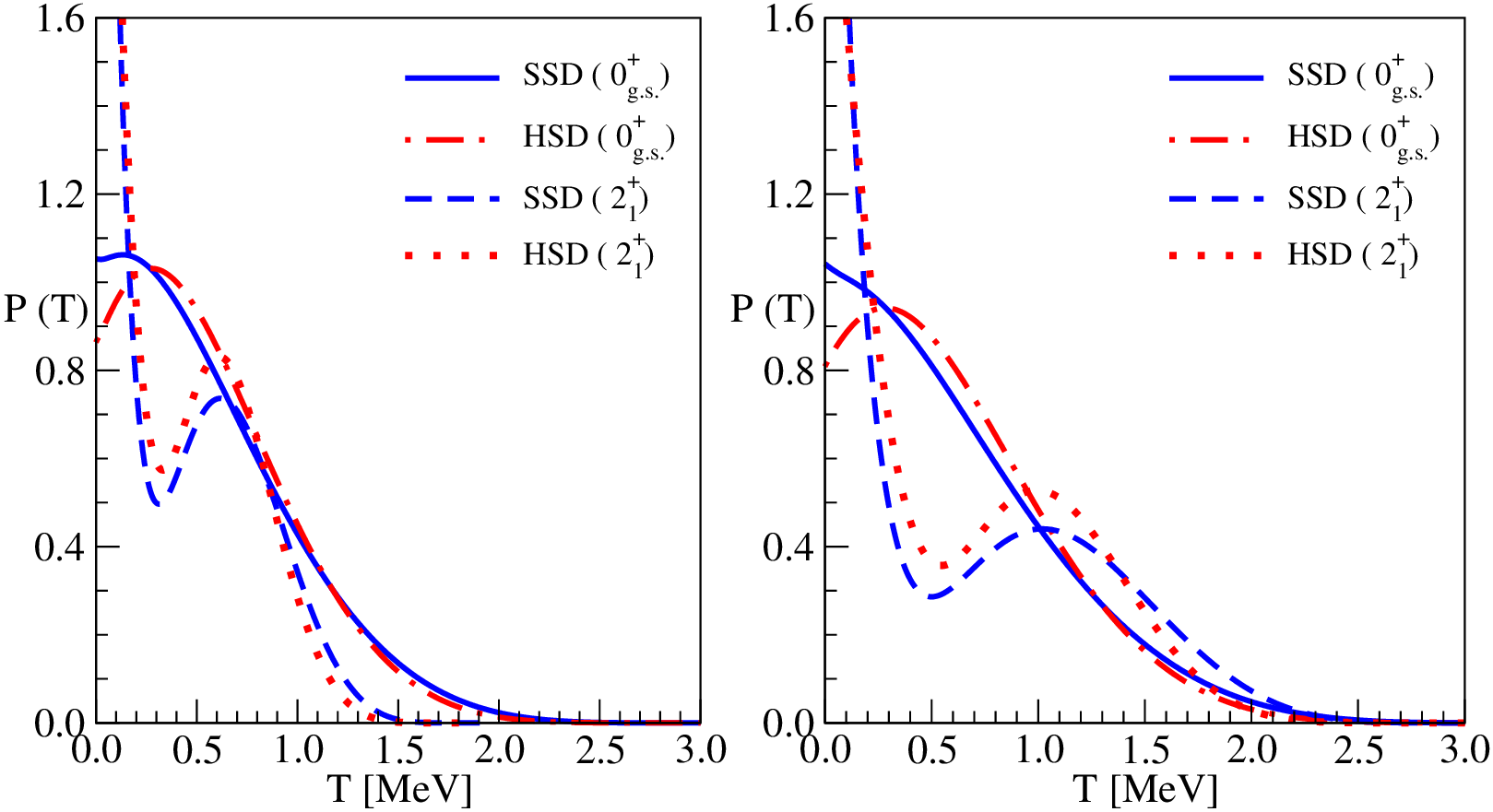}
\caption{The single electron differential decay rate normalized to the total
decay rate vs. the electron energy for $2\nu\beta\beta$-decay of 
$^{100}Mo$ (left panel) and $^{150}Nd$ (right panel) to $0^+$ ground state
and $2^+_1$ excited state. The calculations were performed within the
single-state dominance hypothesis (SSD) and with the assumption of dominance 
of higher lying states (HSD).}
\label{picture1}
\end{figure*}

In Fig. \ref{picture1} we present the single electron differential decay rate 
for the $2\nu\beta\beta$-decay of $^{100}Mo$ and $^{150}Nd$ to $0^+$ ground state
and $2^+_1$ excited state. The SSD results were obtained by assuming the dominance
of  transition through $1^+$ ground state of $^{100}Tc$ and $1^-$ ground state 
of $^{150}Pm$ for $2\nu\beta\beta$-decay of $^{100}Mo$ and $^{150}Nd$, respectively. 
We see that differences between the SSD and HSD predictions are even larger for 
A=150 as for A=100. Unlike there is an unknown $1^+$ low lying state of  $^{150}Pm$ the 
experimental measurement should confirm the HSD results 
for the $2\nu\beta\beta$-decay of $^{150}Nd$. This spectroscopic problem can be 
also addressed by measuring ($d,^2He$) charge exchange  reaction on $^{150}Sm$ 
\cite{frekers}. The recent progress in the field of charge exchange reactions
is encouraging and provide new and sometimes unexpected insight into nuclear
structure phenomena.

\vspace{0.6cm}

\section{The violation of Pauli exclusion principle for neutrinos} 
  
Neutrinos may possibly violate the spin-statistics theorem, and hence obey Bose statistics 
or mixed statistics despite having spin half \cite{ign-kuz}. A violation of the spin-statistics 
relation for neutrinos would lead to a number of observable effects in cosmology and astrophysics. 
In particular, bosonic neutrinos might compose all or a part of the cold cosmological dark matter 
(through bosonic condensate of neutrinos) and simultaneously provide some hot dark matter \cite{dosm}. 
A change of neutrino statistics would have an impact on the evolution of supernovae and on 
the spectra of supernova neutrinos. The Pauli principle violation for neutrinos can be tested 
in the $2\nu\beta\beta$-decay \cite{dosm,Barabash:2007gb}. 

Qualitative features of the $2\nu\beta\beta$-decay in the presence of bosonic neutrinos 
can be understood by the fact that two contributions to the amplitude of the decay 
from diagram with permuted neutrino momenta have relative plus sign instead of minus
in the Fermi-Dirac case. The decay probability is proportional to the bilinear
combinations $K^{f,b}_m L^{f,b}_n$, $K^{f,b}_m L^{f,b}_n$, $L^{f,b}_m L^{f,b}_n$,
where
\begin{eqnarray}
K^{f,b}_m \equiv 
~~~~~~~~~~~~~~~~~~~~~~~
\nonumber\\
\frac{1}
{E_m  - E_i + p_{10} + k_{10}} \pm 
\frac{1}
{E_m  - E_i + p_{20} + k_{20}},
\nonumber\\
L^{f,b}_m \equiv 
~~~~~~~~~~~~~~~~~~~~~~~
\nonumber\\
\frac{1}
{E_m  - E_i + p_{20} + k_{10}} \pm 
\frac{1}
{E_m  - E_i + p_{10} + k_{20}}. 
\label{prop}
\end{eqnarray}
Here, $E_i$, $E_m$, $p_{i0}$ and $k_{i0}$ ($i=1,2$) are the energies of the initial 
nucleus, intermediate nucleus, electrons and neutrinos, respectively.  
We notice a sign difference between the two energy denominators in (\ref{prop}) 
distinguishing the cases of fermionic (f) and bosonic (b) neutrinos. 

The effect of bosonic neutrinos is different for transitions to $0^+$ ground states 
and $2^+$ excited states. It is because the decay rate to $0^+$ state is governed
by the combinations $(K_m+L_m)(K_n+L_n)$ and decay rate to $2^+$ state is
proportional to combinations $(K_m-L_m)(K_n-L_n)$. By
approximating these combinations one finds significantly different expressions
for bosonic and fermionic neutrinos:
\begin{eqnarray}
(K^b_m+L^b_m) &\approx& 
2 \frac{(k_{20} - k_{10})}{( E_m  - E_i + \Delta)^2},
\nonumber\\
(K^b_m - L^b_m) &\approx&  
4 \frac{(p_{20} - p_{10})}{(E_m  - E_i + \Delta)^2},
\nonumber \\
(K^f_m +L^f_m) &\approx&  \frac{4}{E_m  - E_i + \Delta },
\nonumber\\
(K^f_m -L^f_m) &\approx& 2\frac{(p_{20} - p_{10})(k_{20} - k_{10})}
{( E_m - E_i + \Delta )^3}.
\label{KLfb}
\end{eqnarray}
Here, $\Delta$ is the average energy of the leptonic pair.

\begin{figure*}[t]
\includegraphics[height=6.5cm,width=8cm]{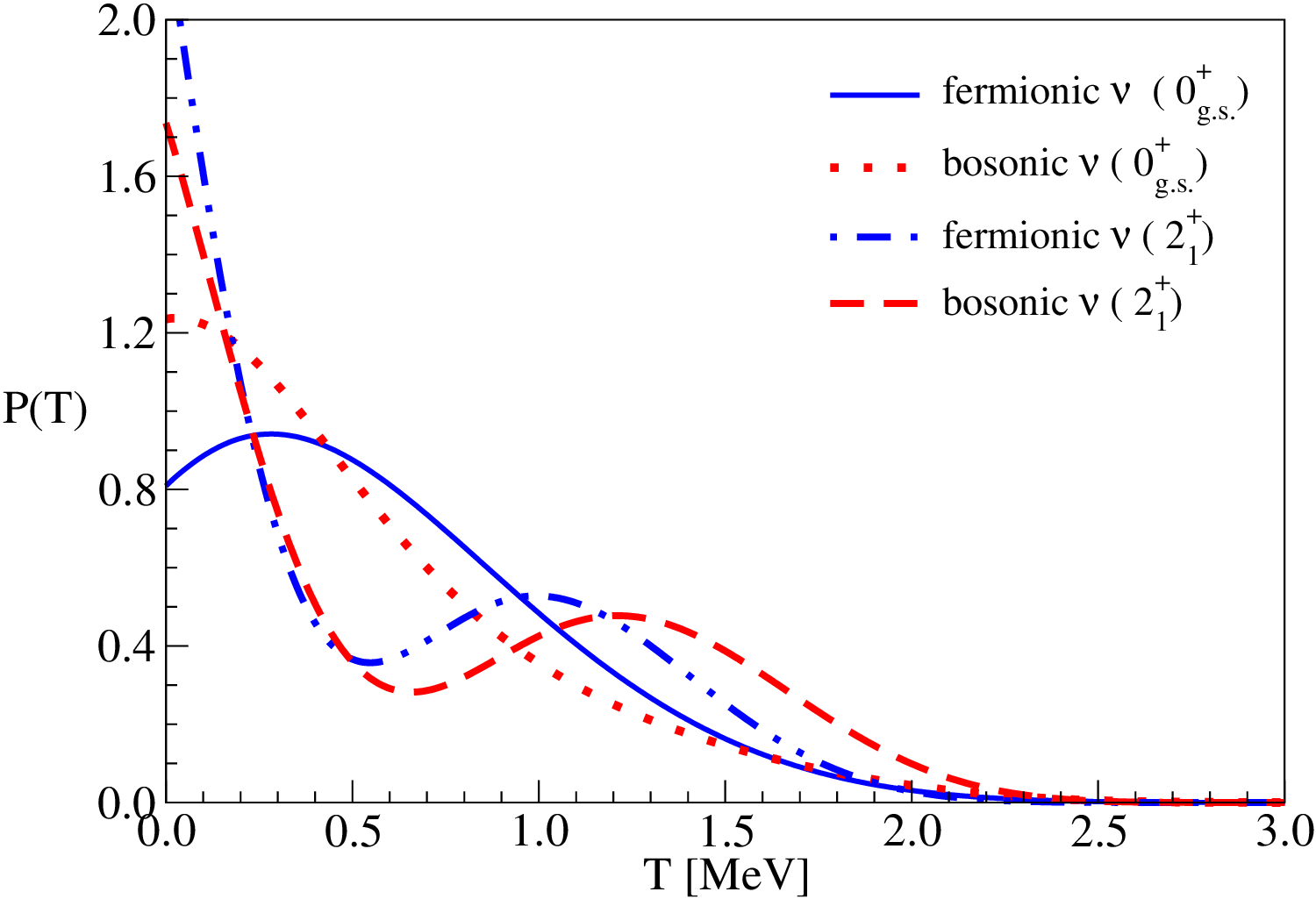}
\hspace*{0.5cm}
\includegraphics[height=6.5cm,width=8cm]{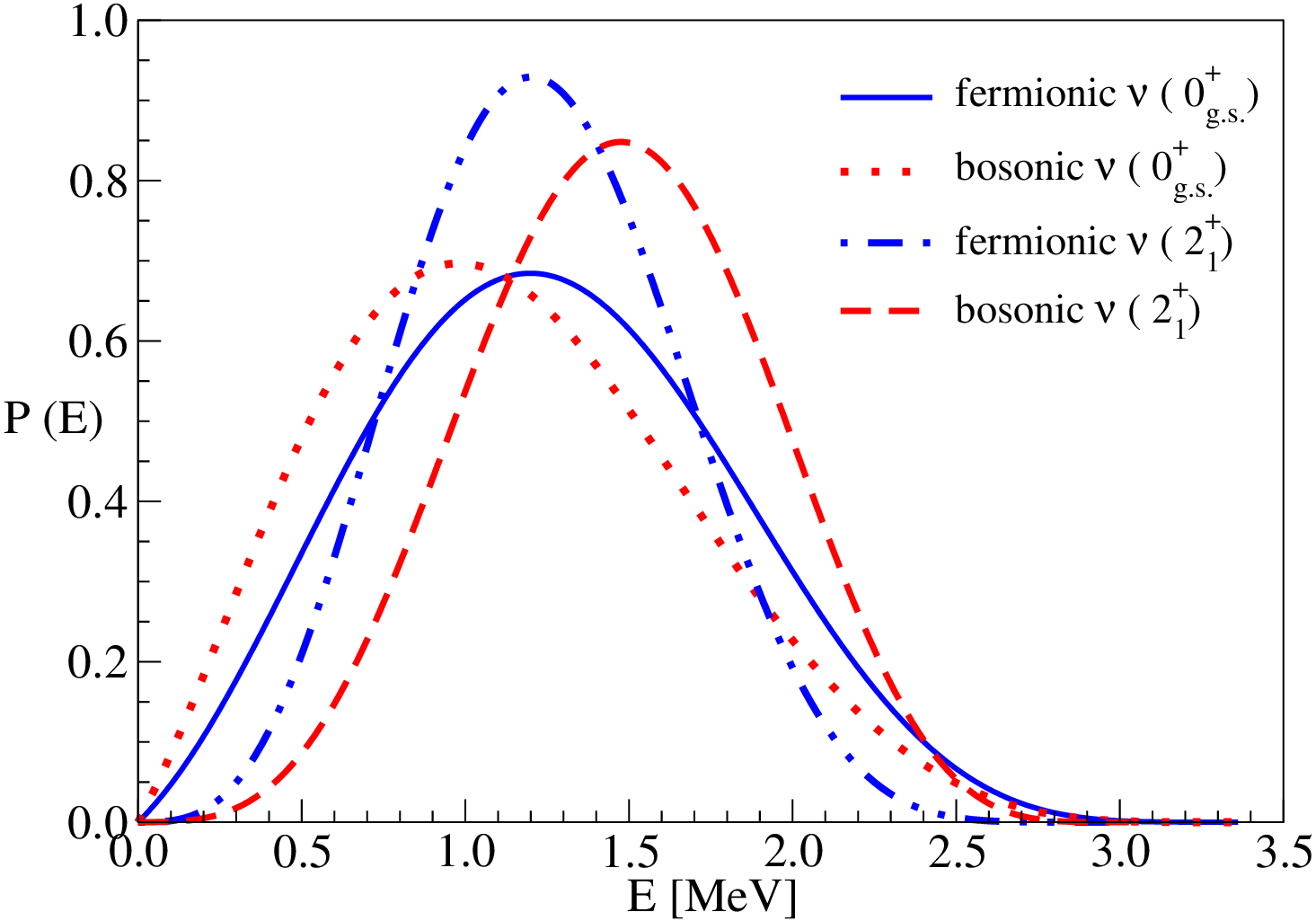}
\caption{The single electron differential rate normalized to the total decay rate vs. 
the electron energy (left panel) and the differential decay rate normalized to the
total decay rate vs. the sum of kinetic energy for $2\nu\beta\beta$-decay of $^{150}Nd$
to the ground $0^+_{g.s.}$ and excited $2^+_1$ states of final nucleus. The results are 
presented for the cases of pure fermionic and pure bosonic neutrinos. The calculations
have been performed with the HSD assumption.}
\label{picture2}	 
\end{figure*}

In Fig. \ref{picture2} we show the energy distributions of outgoing electrons
calculated for the $2\nu\beta\beta$-decay of $^{150}Nd$ to the $0^+_{g.s.}$
ground state and $2^+_1$ excited state with the HSD assumption.
Both for the single electron differential decay rate (left panel) and for differential 
decay rate as function of the sum of kinetic energy of outgoing electrons (right panel) 
the results are significantly different for bosonic and fermionic neutrinos. In particular,
maxima of the distributions for transitions to $0^+_{g.s.}$ and $2^+_1$ states 
(right panel) are about at the same position for fermionic neutrinos and are shifted each 
to other by about 0.5 MeV for bosonic neutrinos.  

\vspace*{0.6cm}

\section{Conclusions}

We offered some arguments that the SSD hypothesis is not expected to be realized 
for $2\nu\beta\beta$-decay of $^{150}Nd$. This might be confirmed within the NEMO
3 and SuperNEMO experiments. However, if there is an unknown low-lying $1^+$
state of $^{150}Pm$ the conclusion might be opposite. This 
issue can be addressed by the (d,$^{2}He$) charge-exchange  experiment on $^{150}Sm$
target \cite{frekers}.

In addition, we showed that a study of the $2\nu\beta\beta$-decay of $^{150}Nd$
can provide a sensitive test of the Pauli exclusion principle and statistics of 
neutrinos. For that purpose an experiment (e.g., SuperNEMO) is needed with a 
precision measurement of differential characteristics for $2\nu\beta\beta$-decay
transitions to $0^+$ ground and $2^+_1$ excited states. 

The work of R. D. and F. \v S. was supported in part 
by the Deutsche Forschungsgemeinschaft (grant No. 436SLK 17/298).
We thank also EU ILIAS project under the Contract No. 
RII3-CT-2004-506222.

\end{document}